\documentclass{aa}  

\usepackage{graphicx}
\usepackage{natbib}
\bibpunct{(}{)}{;}{a}{}{,} 
\usepackage{txfonts}
\begin{document}
   \title{The discovery of diffuse steep spectrum sources in Abell~2256}

   \author{R.~J. van Weeren \inst{1}
          \and H.~T. Intema \inst{1}
          \and J.~B.~R. Oonk \inst{1}
          \and H.~J.~A. R\"ottgering\inst{1}
          \and T.~E. Clarke \inst{2}
          }

   \institute{Leiden Observatory, Leiden University,
              P.O. Box 9513, NL-2300 RA Leiden, The Netherlands\\
              \email{rvweeren@strw.leidenuniv.nl}
         \and Naval Research Laboratory, 4555 Overlook Avenue SW, Washington~D.~C. 20375 and Interferometrics, Inc., 13454 Sunrise Valley
Drive No.~240, Herndon, VA~20171
                }


 
\abstract
    {Hierarchical galaxy formation models indicate that  during their lifetime galaxy clusters undergo several mergers. An example of such a merging cluster is \object{Abell~2256}. Here we report on the discovery of three diffuse radio sources in the periphery of Abell~2256, using the Giant Metrewave Radio Telescope (GMRT). 
    }
   {The aim of the observations was to search for diffuse ultra-steep spectrum radio sources within the galaxy cluster Abell~2256.
   }
   {We have carried out GMRT $325$~MHz radio continuum observations of Abell~2256. V, R and I band images of the cluster were taken with the 4.2m William Herschel Telescope (WHT).
   }
   { We have discovered three diffuse elongated radio sources located about 1~Mpc from the cluster center. Two are located to the west of the cluster center, and one to the southeast. 
   The sources have a measured physical extent of 170, 140 and 240~kpc, respectively. The two western sources are also visible in deep low-resolution $115-165$~MHz Westerbork Synthesis Radio Telescope (WSRT) images,  although they are blended into a single source. For the combined emission of the blended source we find an extreme spectral index ($\alpha$) of $-2.05 \pm 0.14$ between 140 and 351~MHz. The extremely steep spectral index suggests these two sources are most likely the result of adiabatic compression of fossil radio plasma due to merger shocks. For the source to the southeast, we find that ${\alpha < -1.45}$ between 1369 and 325~MHz. We did not find any clear optical counterparts to the radio sources in the WHT~images.
   }
  {The discovery of the steep spectrum sources implies the existence of a population of faint diffuse radio sources in (merging) clusters with such steep spectra that they have gone unnoticed in higher frequency ($\gtrsim 1$~GHz) observations. Simply considering the timescales related to the AGN activity, synchrotron losses, and the  presence of shocks, we find that most massive clusters should possess similar sources.  An exciting possibility therefore is that such sources will determine the general appearance  of clusters in low-frequency high resolution radio maps as will be produced by for example LOFAR or LWA.
   }

   
   \maketitle

\section{Introduction}
Models of large-scale structure (LSS) formation show that nearly all galaxy clusters grow via mergers of smaller clusters and sub-structures. These mergers can create shocks and turbulence within the intra cluster medium (ICM) that are likely to amplify magnetic fields and accelerate relativistic particles. In the presence of magnetic fields, these particles will emit synchrotron radiation, observable at radio frequencies  \citep[see the review by][and references therein]{2008SSRv..134...93F}. Giant peripheral radio relics are thought to be tracers of the shock waves generated by cluster mergers \citep{1998A&A...332..395E, 2000ApJ...542..608M}. At the location of the shock front particles are accelerated via diffusive shock acceleration (DSA) by the Fermi-I process  \cite[e.g.,][]{1983RPPh...46..973D, 1987PhR...154....1B, 1991SSRv...58..259J, 2001RPPh...64..429M}. Shock waves generated by cluster mergers can also compress fossil radio plasma from previous episodes of AGN activity and produce radio relics \citep{2001A&A...366...26E, 2002MNRAS.331.1011E}. Proposed examples of these radio relics, also called radio \emph{phoenices}, are the sources found by \cite{2001AJ....122.1172S}. These sources have a steep spectral index\footnote{$F_{\nu} \propto \nu^{\alpha}$, with $\alpha$ the spectral index} ($\alpha \lesssim-1.5$) and filamentary morphologies. The steep spectral indices are the result of synchrotron and inverse Compton (IC) losses.
\begin{figure*}
    \begin{center}
      \includegraphics[angle = 90, trim =0cm 0cm 0cm 0cm,width=1.0\textwidth]{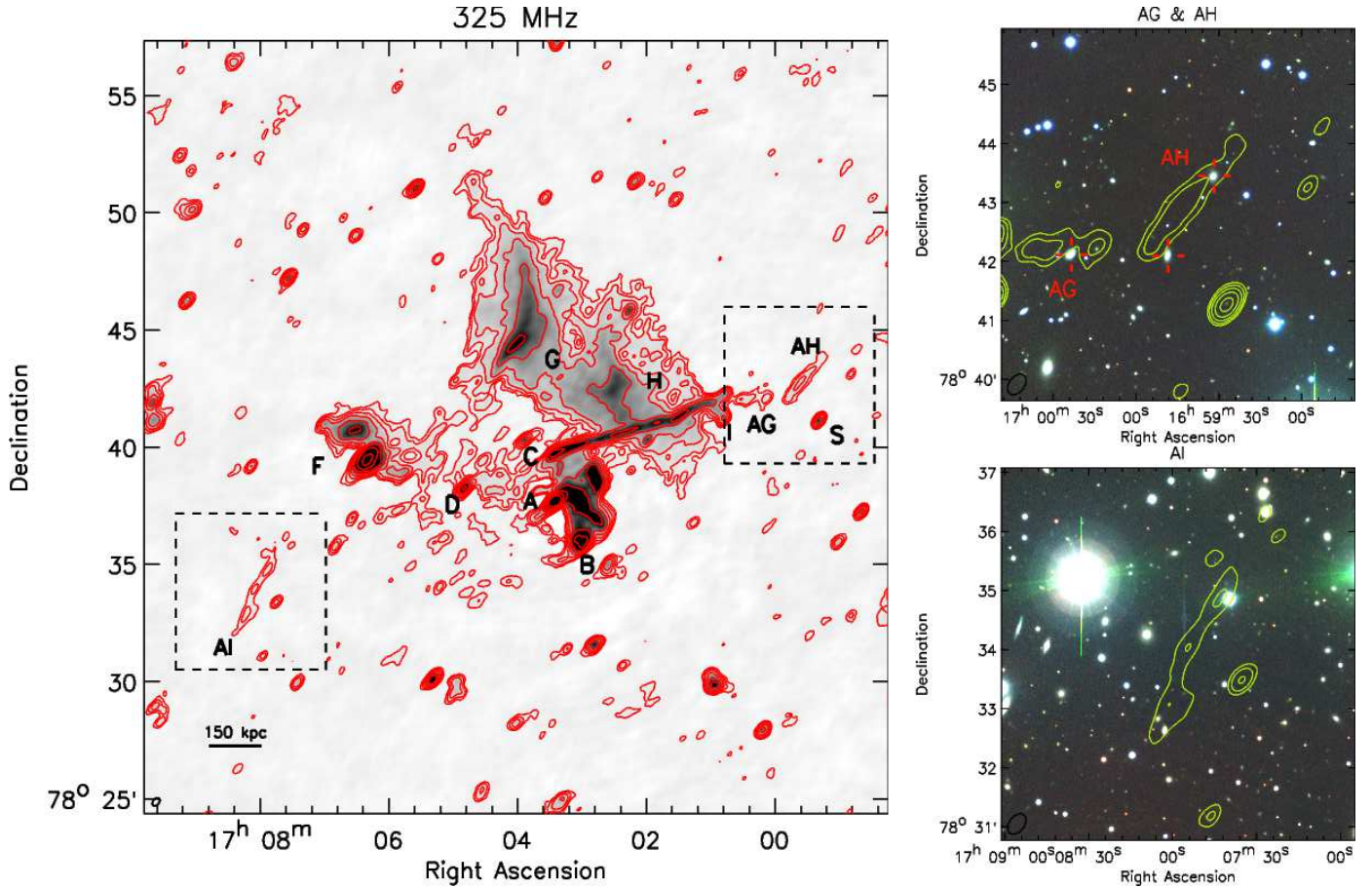}
       \end{center}
      \caption{Left: Radio map at 325 MHz with a restoring beam size of $23.9\arcsec \times 15.9\arcsec$, indicated in the bottom left corner. The red contours are drawn at levels of $[1, 2, 4, 8, \ldots]  \times 3\sigma_{\mathrm{rms}}$~$\mu$Jy~beam$^{-1}$. Various sources are labeled with capital letters. Right: Optical images around AG+AH (top right panel) and AI (bottom right panel) composed out of V (blue), R (green), and I (red) band images. }
                  \label{fig:map325}
       \end{figure*}

\object{Abell~2256} is a massive cluster located at a redshift of $0.0581$ \citep{1999ApJS..125...35S}. Extensive observations from X-ray to radio wavelengths have shown that the cluster is presently undergoing a merger with a smaller sub-structure. These X-ray observations revealed significant sub-structure within the cluster. The most prominent feature are two separate X-ray peaks \citep{1991A&A...246L..10B, 1994Natur.372..439B, 2002ApJ...565..867S} providing strong evidence that Abell~2256 is indeed undergoing a merger. 
At radio wavelengths, the cluster shows a very complex morphology. The cluster is known to host a bright peripheral radio relic, a radio halo, a number of complex relatively compact filamentary sources, and several ``head-tail'' sources \citep{1976A&A....52..107B, 1979A&A....80..201B, 1994ApJ...436..654R, 2003AJ....125.2393M, 2006AJ....131.2900C, 2008A&A...489...69B}.

Here we present deep 325~MHz radio continuum observations of Abell~2256 with the GMRT, complemented by WSRT $115-165$~MHz observations and optical WHT imaging. In this paper we focus on the discovery of three diffuse sources in the cluster periphery. A detailed comparison of the other sources in the cluster has been presented by \cite{intema_phd}. The layout of this paper is as follows. In Sect.~\ref{sec:obs-reduction} we give an overview of the observations and data reduction. In Sect.~\ref{sec:results} we present the radio and optical maps. We end with a discussion and conclusions in Sects.~\ref{sec:discussion} and \ref{sec:conclusion}.
Throughout this paper we assume a $\Lambda$CDM cosmology with $H_{0} = 73$~km~s$^{-1}$~Mpc$^{-1}$, $\Omega_{m} = 0.27$, and $\Omega_{\Lambda} = 0.73$. At the distance of Abell~2256 1\arcsec~corresponds to $1.08$~kpc \citep{2006PASP..118.1711W}.

\section{Observations \& data reduction}
\label{sec:obs-reduction}
Abell~2256 was observed with the GMRT at a frequency of 325~MHz. The observations recorded both LL and RR polarizations in two sub-bands (IFs) with a total bandwidth of 32~MHz, and $128$~frequency channels per IF. The observations were carried out on May~25~\&~26, 2008. The absolute flux scale was set according to the \cite{perleyandtaylor} extension of the \cite{1977A&A....61...99B} scale using the calibrators 3C48, 3C147, and 3C286. The total on-source time for Abell~2256 was about 9.5~hours. The data were reduced using the NRAO Astronomical Image Processing System (AIPS). The data was corrected for the instrumental bandpass and complex gain solutions were determined using the calibrators and applied to our target source. The sidelobes from six bright sources affected our target source in the center of the field. We removed these six sources using the ``peeling''-method \citep[e.g.,][]{2004SPIE.5489..817N}.  We used the polyhedron imaging method  \citep{1989ASPC....6..259P, 1992A&A...261..353C}  to reduce the effects of non-coplanar baselines. 
The final primary beam corrected image was made using robust weighting set to $0.5$ \citep{briggs_phd}, yielding a synthesized beam of $23.9\arcsec\times15.9\arcsec$. The rms noise in this image was 119~$\mu$Jy~beam$^{-1}$. 

A2256 was observed with the WSRT in 8 IFs, each IF having 2.5~MHz bandwidth and 128 channels, placed between 115 and 165~MHz. The total integration time was $6 \times12$~hours using six different configurations (with four moveable antennas) to suppress grating lobes. Each 12~hour observation and IF were calibrated separately, correcting for the bandpass and slow gain variations (using 3C48 and 3C295). The data was further self-calibrated using three rounds of phase and one round of amplitude and phase selfcalibration. The data for the six configurations were then combined per IF and imaged separately. The noise in the images ranged between $5.0-14.6$~mJy~beam$^{-1}$, and the resolution varied from 175\arcsec~at 115~MHz to 130\arcsec~at 165~MHz. For more details on the reduction see \cite{intema_phd}.

Optical V, R and I band images were taken at the location of our newly discovered sources with the 4.2m WHT telescope (PFIP CCD camera) on April 19,~2009.  The seeing during the observations was about 1.5\arcsec, the total exposure time was 600~s per band. The images were reduced with IRAF \citep{1986SPIE..627..733T, 1993ASPC...52..173T} using the \emph{mscred} package \citep{1998ASPC..145...53V}. All images were flat fielded and bias-corrected. The I band image was fringe corrected. Zero-points were determined using various observations of standard stars taken during the night. The final images have a depth (SNR of 4 for point sources) of approximately $23.9$, $24.2$, $22.8$ magnitude (Vega) in the V, R and I band, respectively.
\section{Results}
\label{sec:results}

       We adopt the radio source labeling from \cite{1976A&A....52..107B} and \cite{1994ApJ...436..654R}. The 325~MHz GMRT image (Fig.~\ref{fig:map325}) clearly shows the presence of the known AGN related sources (A, B, C, and I), the bright peripheral relic (the combined emission from G and H) and the complex source F. The radio halo is only partly visible around source D since the shortest baselines were severely affected by RFI and had to be flagged. In this paper we focus on three new diffuse sources discovered in the maps, which we label source AG, AH, and AI, see Fig.~\ref{fig:map325}. AG and AH are neighboring sources (about 1\arcmin) to the west of the cluster center and have an extent of 130\arcsec~and 160\arcsec, respectively. Source AI is located on the SE side of the cluster and has an extent of 220\arcsec. The integrated 325~MHz fluxes are $7 \pm 1$,  $12\pm 1$, and $11 \pm 1$~mJy for sources AG, AH and AI, respectively (by summing the flux within a region half an arcminute outside the $3\sigma$ contours). Sources AG and AH are also seen in the 351~MHz image presented by \cite{2008A&A...489...69B}. However, due to the lower resolution of this image, the diffuse nature of the sources is not apparent. Source AI is also seen in the same image, but only the southern brighter part is visible. The northern half of AI is too faint to be visible in this map. At a redshift of $0.0581$ the sources have a physical extent of 140 (AG), 170 (AH) and 240~kpc (AI). They are located at a projected distance of about one Mpc from the cluster center.

Optical images at the position of the radio sources are also shown in Fig~\ref{fig:map325}. At a redshift of 0.0581 any cluster member capable of hosting an AGN should be visible in our optical images. We cannot unambiguously identify an optical counterpart for sources AH and AI. The western part of source AG may be identified with a background galaxy, (most likely) located behind the cluster. However, no radio source at this location is found in 1.4~GHz Very Large Array (VLA) images  \citep{2006AJ....131.2900C}, indicating the source must have a steep spectrum.

\subsection{Spectral indices}
\label{sec:spectralindex}
\begin{figure}
   \begin{center}
    \includegraphics[angle = 90, trim =0cm 0cm 0cm 0cm,width=0.5\textwidth]{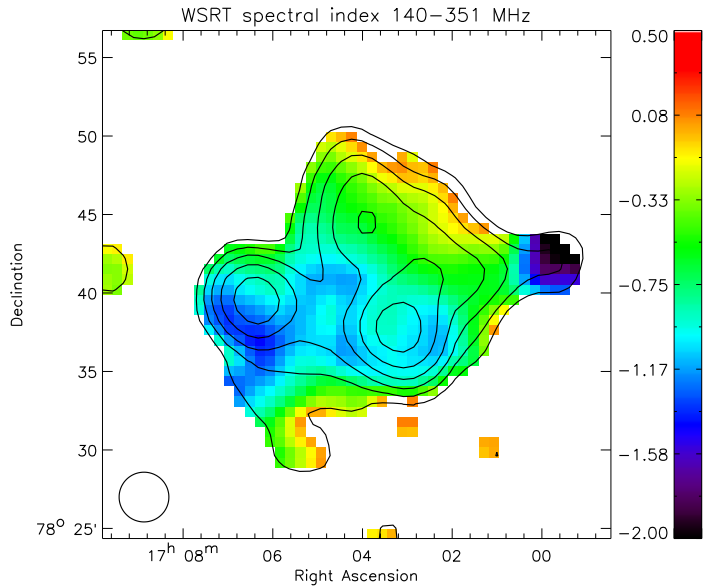}
     \end{center}
    \caption{WSRT spectral index map between 140 and 351 MHz. Total intensity contours at 140 MHz are shown at levels of  $[1, 2, 4, 8, 16, 32,  ...]~\times 30$~mJy~beam$^{-1}$. The total intensity map is a linear combination of six IFs between 115 and 165 MHz. The 351~MHz image comes from \cite{2008A&A...489...69B}.}
    \label{fig:spix_wsrt}
\end{figure}

Sources AG+AH are detected in each individual WSRT band. Source AI is probably detected ($\sim$~$3\sigma$) in a deep low-resolution~(175\arcsec) WSRT~$140$~MHz continuum image, a combination of images from six IFs convolved to 175\arcsec~resolution, see Fig.~\ref{fig:spix_wsrt} (the images of two IFs affected by strong RFI have been left out).  By fitting a single spectral index through the 6 individual IFs, it was found that the spectral index around AG+AH was steeper by about -1.5 units than that of the main relic to the east. A better spectral index map was created using the 351~MHz image from \citeauthor{2008A&A...489...69B} (Fig. 2 convolved to a resolution of 175\arcsec). The spectral index of AG+AH in this map is $-2.05\pm0.14$.

Sources AG, AH and AI are not detected in deep 1369, 1413, 1513, and 1703~MHz VLA C-array (L-band) observations from \cite{2006AJ....131.2900C},  which have noise levels of 34, 40, 40, and 43~$\mu$Jy~beam$^{-1}$, respectively. This implies a steep spectrum for these sources and rules out an identification as a radio-loud AGN within the cluster (which have a typical spectral index of $-0.7$). We combined these four L-band images to create a deeper image, using the 1369~MHz image as a reference. The images were all convolved to a beam size of $16\arcsec~\times~16\arcsec$. The images were scaled to a common flux scale by adopting different spectral indices ranging between $-0.5$ and $-3.0$, but we did not detect the diffuse sources in any of these combined maps. Using a spectral index scaling of $-1.0$ resulted in a noise level about a factor $0.7$ lower compared to the 1369~MHz image. We have taken this image to derive an upper limit on the spectral index between 1369 and 325~MHz. The local rms noise level near AI in the combined L-band image is 58~$\mu$Jy~beam$^{-1}$. AI covers an area of 61 beams (using the same area over which we measured the integrated flux as reported in Sect.~\ref{sec:results}). Assuming that the noise scales $\propto \sqrt{N_{\mathrm{beams}}}$, with $N_{\mathrm{beams}}$ the number of beams, this gives an upper limit for the 1369~MHz flux of 1.36~mJy, using a signal to noise ratio of (SNR) of 3 for a detection. Combined with the 325~MHz flux this implies that $\alpha < -1.45$ for AI (between 325 and 1369~MHz). We do a combined analysis for sources AG and AH as they are not separable in the spectral index map. The rms noise level near AG+AH is 40~$\mu$Jy~beam$^{-1}$. The sources cover an area of 92 beams, which gives an upper limit of 1.15~mJy for the 1369~MHz flux, implying that $\alpha < -1.95$ (between 1369 and 1369~MHz). This is consistent with the spectral index of $-2.05\pm0.14$ between 140 and 351~MHz.

\section{Discussion}

\label{sec:discussion}
 
Three different origins have been proposed for the presence of radio relics in clusters: (i) in-situ acceleration by the DSA mechanism, (ii) adiabatic compression of fossil radio plasma (producing radio \emph{phoenices}), and (iii) \emph{AGN relics}. The first mechanism is thought to be responsible for the presence of giant Mpc-size radio relics in the periphery of clusters, as is the case for the well known bright relic in A2256. In the second case a passing shock wave in the ICM adiabatically compresses fossil radio plasma, producing a radio phoenix. The morphologies of such sources can be filamentary or toroidal \citep{2002MNRAS.331.1011E}. This radio plasma is proposed to originate from a previous episode of AGN activity. After the electrons have lost most of their energy they become invisible as ``radio ghosts''. At the passage of a shock wave they are revived (their brightness increases) and they become visible again with very steep radio spectra. The size of these sources is in general limited to a few hundred kpc as the time to compress larger ghosts would remove most of the energetic electrons responsible for the radio emission by radiative energy losses \citep{2006AJ....131.2900C}. Radio plasma which originated from a previous episode of AGN activity can remain visible as a steep-spectrum AGN relic. These sources are generally fainter than radio phoenices of the same age, as no compression of radio plasma has taken place \citep{2001A&A...366...26E, 2002MNRAS.331.1011E}. These AGN relics do not require shocks from cluster mergers. 

The two radio sources to the west of the cluster center are probably radio phoenices or AGN relics, as they have steep radio spectra, relatively small sizes, are not clearly identified with galaxies, and diffuse in nature. Since the cluster is undergoing a merger, and the sources do not show a symmetric double lobe structure (like some of the AGN relics found by \citet{2009ApJ...698L.163D}), these sources are more likely to be radio phoenices than AGN relics. Spectral modeling as carried out by e.g., \cite{intema_phd} and \cite{2009ApJ...698L.163D} can be used to separate between these two scenarios. At present we do not have enough flux measurements for our sources available to carry out such an analysis.  Source AI is located at the edge of the radio halo, this source could be a radio phoenix or a faint radio relic tracing a shock front.

Since the plasma from both radio phoenices and AGN relics should originate from AGN, we have searched for any nearby galaxies that could have harbored an AGN. These galaxies are marked with red crosses in Fig.~\ref{fig:map325}. Sources AG and AH are located at the end of head-tail source C. It could be that the radio plasma originated from this tail. Other galaxies located in the cluster are also present in the neighborhood: \object{2MASX~J17002352+7842076} \citep[$z=0.058834$, ][]{2002AJ....123.2261B}, is located at the position of AG. For AH we find two possible galaxies, \object{Abell~2256:[BLC2002]~0056} ($z=0.056518$) to the north, and \object{2MASX~J16594813+7842055} ($z=0.055357$) to the south. However, none of these are classified as a giant ellipticals which are more likely to have been active. A typical maximum age for a radio phoenix is $10^8$~yr \citep{2001A&A...366...26E}. Using the velocity dispersion of 1269~km~s$^{-1}$ \citep{2003AJ....125.2393M} for A2256, the displacement radius would be around 130~kpc ($\sim 2\arcmin$). This radius could be several times larger if the radio plasma is older, or if the host galaxy has a higher peculiar velocity. Within the range of several hundreds of kpc many cluster members are found, we therefore cannot identify the sources of the radio plasma.

There are about 50 elliptical galaxies within the cluster which can host an AGN, 15 of which are observed to be radio-loud \citep{2003AJ....125.2393M, 2002AJ....123.2261B}. This radio-loud fraction is consistent with the results from \cite{2005MNRAS.362...25B} for the general population of massive galaxies. 
\cite{2001A&A...366...26E} showed that the fossil radio plasma in radio ghosts needs to be revived within $10^8$ to $10^9$~yr to become a radio phoenix, depending on the synchrotron losses. Here we adopt the more conservative value of $10^8$~yr. The number of radio phoenices in a cluster ($N_{\mathrm{phoenix}}$) is the product of the number of inactive AGNs in a cluster, that produced radio ghosts less than $10^{8}$~yr old ($N_{\mathrm{ghost}}$), and the fraction of them being compressed ($f_{\mathrm{compressed}}$) by (merger) shocks waves: $N_{\mathrm{phoenix}} = N_{\mathrm{ghost}} \times f_{\mathrm{compressed}}$. 
To estimate $N_{\mathrm{ghost}}$ one needs to know which fraction of elliptical galaxies in a cluster goes through phases of AGN activity and their duty cycle. These numbers are not well constrained, but we know that the fraction of active AGN is at least $15/50=0.3$ in the case of A2256. Estimates for the duty cycle of AGN are around $10^8$~yrs \citep[e.g.,][]{2004ApJ...607..800B, 2006MNRAS.367.1121P, 2008MNRAS.388..625S}, but these estimates vary by an order of magnitude and are only for cD galaxies in the center of cool core clusters. Although it is difficult to estimate the number of radio phoenices in a cluster, the fact that there are several radio phoenices in A2256 makes it likely that other merging clusters have radio ghosts, which have been revived by merger shock waves and are visible as radio phoenices.  

\section{Conclusions}

\label{sec:conclusion}

We have presented GMRT 325~MHz radio observations of the cluster Abell~2256. Our radio maps reveal the presence of three diffuse sources located in the periphery of the cluster. The sources have a steep spectrum. For the combined emission from AG and AH we find $\alpha= -2.05 \pm 0.14$ between 140 and 351~MHz. For AI we set an upper limit of $\alpha <-1.45$ between 325 and 1369~MHz. Optical images do not reveal any obvious counterparts. Source AI could be a faint radio relic tracing a shock front in the ICM or a radio phoenix. Detailed spectral index modeling and/or deep X-ray observations (searching for shock fronts within the ICM) could be used to separate between these scenarios. Sources AG \& AH are most likely radio phoenices, fossil radio plasma from a previous episode of AGN activity that has been compressed by merger shock waves. Although, we cannot exclude the possibility that the sources are AGN relics. In this case the sources might be related to the long head-tail source C within the cluster.

Most massive merging clusters should possess similar sources, considering the timescales related to AGN activity, the number of active AGN in clusters, the presence of shocks, and synchrotron and IC losses. These sources are very faint at high-frequencies ($\gtrsim 1$~GHz) so they have gone previously unnoticed. Since the radio spectra of these sources are steep, they will probably determine the observed appearance of clusters in low-frequency radio maps as will be produced for example by LOFAR \citep{2006astro.ph.10596R} or the Long Wavelength Array \citep[LWA;][]{2008ASPC..395..368K}.

\begin{acknowledgements}

We thank the anonymous referee for useful comments. We would like to thank Michiel Brentjens for providing the 351~MHz WSRT map. We thank the staff of the GMRT who have made these observations possible. GMRT is run by the National Centre for Radio Astrophysics of the Tata Institute of Fundamental Research. The Westerbork Synthesis Radio Telescope is operated by the ASTRON (Netherlands Institute for Radio Astronomy) with support from the Netherlands Foundation for Scientific Research (NWO). The William Herschel Telescope is operated on the island of La Palma by the Isaac Newton Group in the Spanish Observatorio del Roque de los Muchachos of the Instituto de Astrof'sica de Canarias. Basic research in radio astronomy at the NRL is supported by 6.1 Base funding. RJvW acknowledges funding from the Royal Netherlands Academy of Arts and Sciences.

\end{acknowledgements}

\bibliographystyle{aa}
\bibliography{12934b}
\end{document}